\documentclass[9pt,twocolumn,twoside]{osajnl}
\journal{ao} % Choose journal (ao,jocn,josaa,josab,ol,optica,pr)

\setboolean{shortarticle}{false}
\usepackage{lineno}
\title{Space-variant Shack-Hartmann wavefront sensing based on affine transformation estimation}

\author[1,11, $ \dagger $ ]{Fan Feng}
\author[2, $ \dagger $ ]{Chen Liang}
\author[3]{Dongdong Chen}
\author[4]{Ke Du}
\author[2]{Runjia Yang}
\author[5]{Chang Lu}
\author[3]{Shumin Chen}
\author[4,6,7]{Liangyi Chen}
\author[1,8]{Louis Tao}
\author[2,9,10]{Heng Mao}

\affil[1]{Center for Bioinformatics, National Laboratory of Protein Engineering and Plant Genetic Engineering, School of Life Sciences, Peking University, Beijing 100871, China}
\affil[2]{LMAM, School of Mathematical Sciences, Peking University, Beijing 100871, China}
\affil[3]{School of Software and Microelectronics, Peking University, Beijing 100871, China}
\affil[4]{State Key Laboratory of Membrane Biology, Beijing Key Laboratory of Cardiometabolic Molecular Medicine, Institute of Molecular Medicine, Center for Life Sciences, College of Future Technology, Peking University, Beijing 100871, China}
\affil[5]{School of Instrumentation and Optoelectronic Engineering, Beihang University, Beijing 100191, China}
\affil[6]{PKU-IDG/McGovern Institute for Brain Research, Beijing 100871, China}
\affil[7]{Beijing Academy of Artificial Intelligence, Beijing 100871, China}
\affil[8]{Center for Quantitative Biology, Peking University, Beijing 100871, China}
\affil[9]{Beijing Advanced Innovation Center for Imaging Theory and Technology, Capital Normal University, Beijing 100871, China}
\affil[10]{heng.mao@pku.edu.cn}
\affil[11]{ffeng@pku.edu.cn}
\affil[ $ \dagger $ ]{These authors contributed equally to this work}

\begin{abstract}
The space-variant wavefront reconstruction problem inherently exists in deep tissue imaging. In this paper, we propose a framework of Shack-Hartmann wavefront space-variant sensing with extended source illumination. The space-variant wavefront is modeled as a four-dimensional function where two dimensions are in the spatial domain and two in the Fourier domain with \textit{priors} that both gently vary. Here, the affine transformation is used to characterize the wavefront space-variant function. Correspondingly, the zonal and modal methods are both escalated to adapt to four-dimensional representation and reconstruction. Experiments and simulations show double to quadruple improvements in space-variant wavefront reconstruction accuracy compared to the conventional space-invariant correlation method.
\end{abstract}

\setboolean{displaycopyright}{true}

\begin{document}

\maketitle

\section{Introduction}
The space-variant problem in deep tissue imaging has bothered biologists and microscopists for decades. Various methods have been developed to resolve the problem and are roughly categorized into two types: isotropic approximation and the conjugate method. The isotropic approximation method, which assumes an isotropic zone within a certain range to be space-invariant, has been used in laser point scanning microscopy \cite{Wang2014}, selected plane illumination microscopy \cite{Liu2018}, structured illumination microscopy (SIM) \cite{Hoffman2020}, Fourier ptychography microscopy (FPM) \cite{Ou2014}, etc. However, the isotropic zone is usually small; thus, this method is not efficient for large field of view (FOV) applications that take numerous snapshots for a single image. The counterpart conjugate method assumes that the wavefront fluctuations originate from one plane near the sample, instead of the pupil plane, as used in widefield microscopy \cite{Li2015,Mertz2015}. Alternatively, the pupil wavefronts from different FOVs are separated in space and then sensed and corrected, as used in laser point scanning microscopy \cite{Park2017}. However, the sample-induced wavefront is hardly simply modeled by a phase screen. Otherwise, the images may not be fully restored due to this model error. In addition, sample-induced aberration could be modeled by multiple planes, as used in astronomy and called multiconjugate adaptive optics \cite{Monty2021}, which could definitely improve the system performance but also dramatically increase the complexity and cost of the imaging system; thus, this technology has not yet been used in microscopy. 

The space-variant imaging problem can be divided into two steps: wavefront sensing and image restoration. This paper focus on how to model and reconstruct space-variant wavefront. For direct wavefront measurement, the Shack-Hartmann wavefront sensor (SHWFS) is the most popular way. The SHWFS consists of a microlens array and a camera, and the camera is usually located on the back focal plane of the microlens. The subspot images on the camera reveal the gradient of the incident wavefront by its centroid drifts within the corresponding lenslet area under the point source illumination as a guide star. However, this conventional Shack-Hartmann (SH) scheme only represents the wavefront at one designated field of angle, that is, the angle of the guide star \cite{Feng2018}. Besides, the averaged wavefront across the isotropic zone is represented by the local shifts of the optical spot images based on the two-photon laser descanning technique \cite{Wang2014}. In addition, the averaged wavefront across the entire FOV can also be calculated by the correlation between the subimages with and without wavefront aberrations under extended source illumination \cite{Woger2009,Honma2019}. Similar to obtaining the averaged wavefront, the difference is that the descanning technique adds all intensities of the subimages to improve the signal-to-noise ratio since the fluorescence is very weak, while the correlation method uses the subimages themselves. However, all these Shack-Hartmann schemes cannot present a space-variant wavefront.

In this paper, we illustrate an SHWFS scheme to reconstruct a space-variant wavefront with extended source illumination. The affine transformation is used to evaluate subimage transformation across the FOV which implies that the wavefront changes slowly. Furthermore, the zonal and modal methods that reconstruct the wavefront from the gradient map are both upgraded to meet the requirement for space-variant wavefront reconstruction. In particular, a dual-orthogonal model is proposed to represent the space-variant wavefront function, and the relation between the affine coefficients and the dual-orthogonal coefficients is derived and used in the modal reconstruction. Numerical simulations and imaging experiments are both implemented.

\section{Theory}
\label{sec:Theory}

\subsection{Space-variant wavefront function modal}
\label{subsec:Space-variant wavefront function modal}
First, let $ \phi \left( \xi ,\eta \right) $ denote the wavefront on a circular pupil in the optical system while moderate enough to be well represented by a sum of orthogonal polynomials

\begin{equation}
 \phi \left( \xi ,\eta \right)=\sum\limits_{p=1}^{P}{{{c}_{p}}{{R}_{p}}\left( \xi ,\eta \right)},
\label{eq:phi}
\end{equation}
where $ \left( \xi ,\eta \right) $ are the global normalized lateral coordinates of the pupil plane, or Fourier domain, $ {{R}_{p}}\left( \xi ,\eta \right) $ are the $ p  $th orthogonal polynomials $ {{c}_{p}} $ and are the corresponding coefficients, e.g., the Zernike polynomials $ {{Z}_{p}}\left( \xi ,\eta \right) $ for the circular domain or the Legendre polynomials $ {{L}_{p}}\left( \xi ,\eta \right) $ for the rectangular domain, $P$ is the maximum representation term for the pupil plane.

\begin{figure}[htbp!]
\centering
\includegraphics[width=\linewidth]{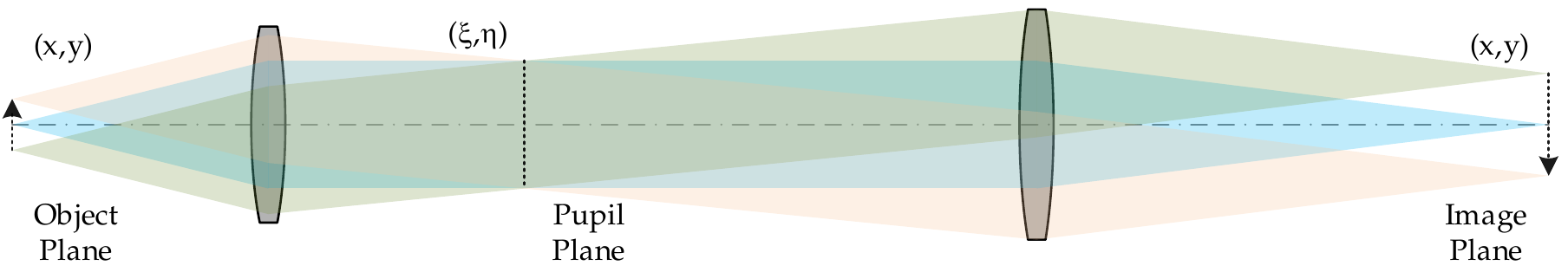}
\caption{Coordinates defined in the imaging system.}
\label{fig:Coordinates system}
\end{figure}

In the case of space-variant wavefront sensing, such as in deep-tissue imaging, the wavefront varies along different positions. Mathematically, the constant coefficients $ c_p $ in \eqref{eq:phi} evolve into the space-variant function $ c_p(x,y) $ , where $ (x,y) $ is the lateral coordinates of the object and image plane, also normalized to 1 for orthogonal representation, as shown in Fig. \ref{fig:Coordinates system}; thus the magnification and the inversion of coordinates on object and image planes are neglected. 

Therefore, we define the space-variant wavefront, 

\begin{equation}
	\phi \left( x,y,\xi ,\eta \right)=\sum\limits_{p=1}^{P}{{{c}_{p}}\left( x,y \right){{R}_{p}}\left( \xi ,\eta \right)}.
\label{eq:space-variant wavefront}
\end{equation}

Similarly, we assume that the space-variant function $ c_p(x,y) $ varies slowly; thus, the dual-orthogonal modal of the space-variant wavefront function is given as

\begin{equation}
 \phi \left( x,y,\xi ,\eta \right)=\sum\limits_{q=1}^{Q}{\sum\limits_{p=1}^{P}{{{e}_{p,q}}{{T}_{q}}\left( x,y \right){{R}_{p}}\left( \xi ,\eta \right)}},
\label{eq:dual-orthogonal space-variant wavefront}
\end{equation}
where
\begin{equation}
 {{c}_{p}}\left( x,y \right)=\sum\limits_{q=1}^{Q}{{{e}_{p,q}}{{T}_{q}}\left( x,y \right)},
\label{eq:orthogonal space-variant function}
\end{equation}
and $ {{T}_{q}}\left( x,y \right) $ are the $ q$th orthogonal polynomials,  $ {{e}_{p,q}} $ are the corresponding coefficients, and $Q$ is the maximum representation term for the pupil plane.. The purpose of our work is to reconstruct the space-variant wavefront $ \phi \left( x,y,\xi ,\eta \right) $ by retrieving the dual-orthogonal coefficients $ {{e}_{p,q}} $ , as the modal method, or discrete $ \phi \left( x_m,y_n,\xi_p ,\eta_q \right) $ on position $ \left( x_m,y_n,\xi_p ,\eta_q \right) $ , as the zonal method.

\subsection{Affine transformation estimation for evaluating the space-variant function in Shack-Hartmann subapertures}
\label{subsec:Affine transformation estimation for evaluating space-variant function in Shack-Hartmann subapertures}

\begin{figure}[htbp!]
\centering
\includegraphics[width=\linewidth]{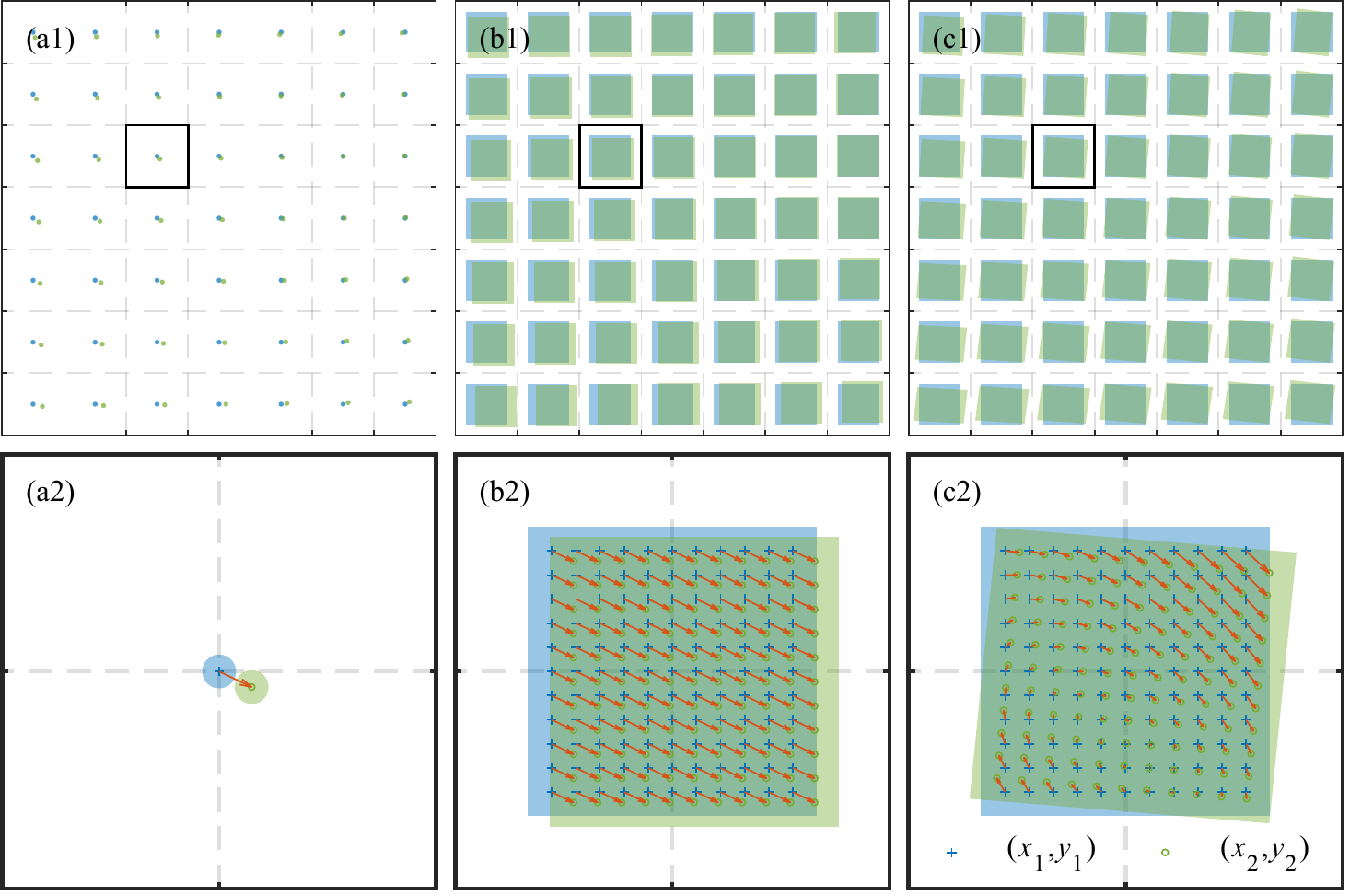}
\caption{Schematic of wavefront sensing in Shack-Hartmann sensor. Space-invariant wavefront sensing with (a1) point and (b1) extended source illumination. (c1) Space-variant wavefront sensing with extended source illumination. Local wavefront gradients represented by (a2) subspot shift, (b2) subimage shift, and (c2) subimage transformation. (a2-c2) Subimages zoom of sections in the box of (a1-c1).}
\label{fig:Schematic}
\end{figure}

\begin{figure*}[htbp!]
\centering
\includegraphics[width=12cm]{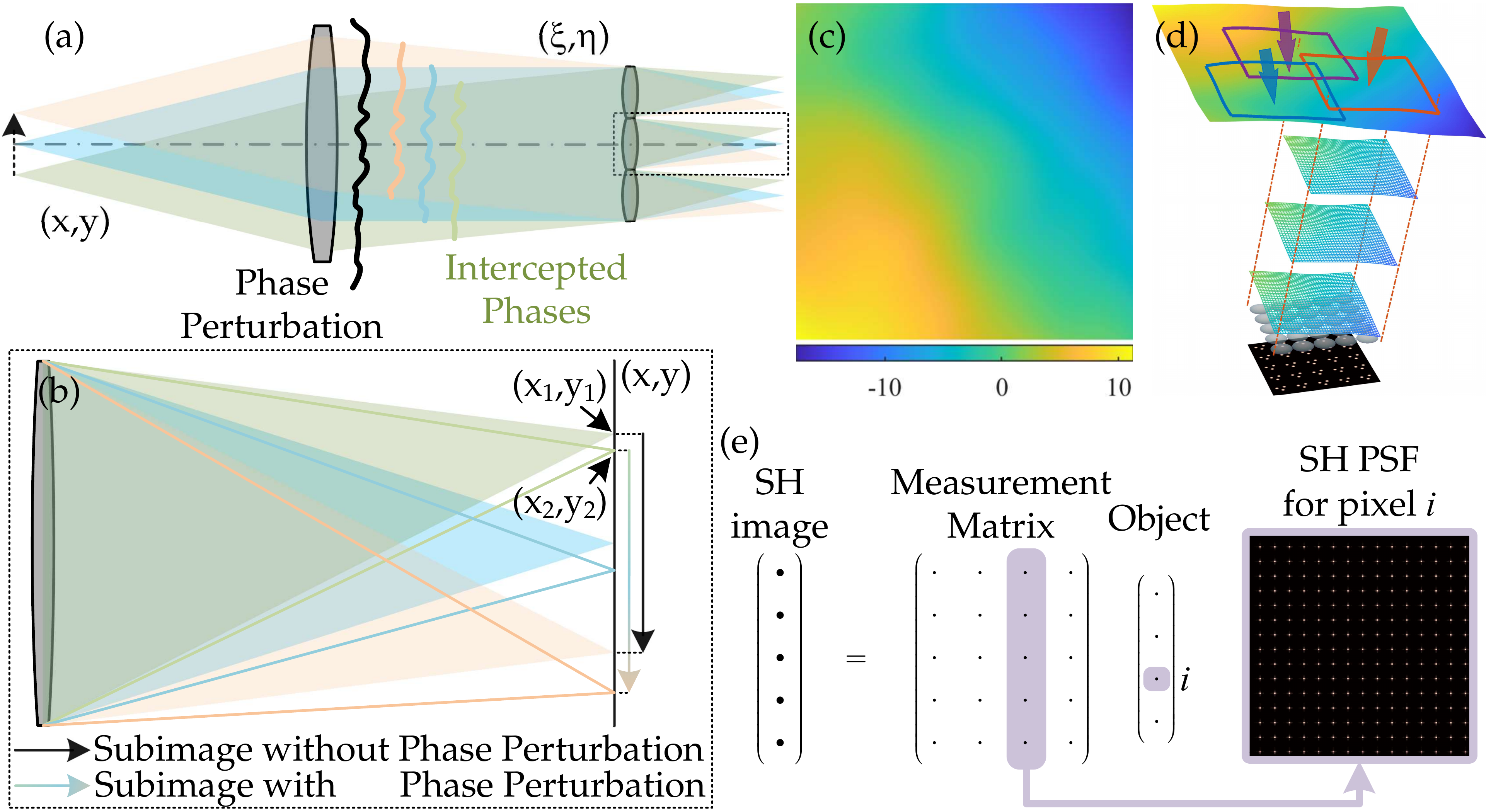}
\caption{Simulations on Shack-Hartmann space-variant wavefront sensing system and geometrical projection schematic for generating wavefronts with different views. (a) SH system with phase screen perturbation, (b) subimages formation in a single lenslet with and without phase perturbation, (c) phase screen distribution used in simulations, scaled in radians, (d) 3D rendering for geometrical projection schematic that the phase screen projects onto the microlens array. (e) Forward process of SH imaging by using the measurement matrix system representation.}
\label{fig:Geometrical projection}
\end{figure*}

In SH, the light field projects onto a microlens array and forges an image on the back focal plane with a subimage array, as shown in Fig. \ref{fig:Schematic}(a1-c1). For the scenario of the conventional space-invariant wavefront, only one gradient vector is obtained from subimages in one lenslet, as shown in Fig. \ref{fig:Schematic}(a2-b2). The centroid shift or correlation algorithm, depending on which light source is used, the point or extended, can be used for representing the local slope. On the other hand, for sensing space-variant wavefront, the subimages deform as shown in Fig. \ref{fig:Schematic}(c2), rather than just translation, as shown in Fig. \ref{fig:Schematic}(b2). We used the affine transformation to characterize the deformation of the subimages in one lenslet

\begin{equation}
\left[\begin{array}{c}
x_2-x_1 \\
y_2-y_1 \\
1
\end{array}\right]=\left[\begin{array}{ccc}
a_1 & a_2 & a_3 \\
b_1 & b_2 & b_3 \\
0 & 0 & 1
\end{array}\right]\left[\begin{array}{c}
x_1 \\
y_1 \\
1
\end{array}\right],
\label{eq:affine transformation}
\end{equation}
where $ (x_1,y_1) $ is an arbitrary point on the $ (x,y) $ plane of the corresponding lenslet area since the back focal plane of the lenslet is also the image plane,  $ (x_2,y_2) $ is the coordinate after deformation, as shown in Fig. \ref{fig:Schematic}(c2) and Fig. \ref{fig:Geometrical projection}(a,b), $ [x_2-x_1,y_2-y_1]^T $ presents the centroid shift of point $ (x_1,y_1) $ , and $ \{a_1,a_2,a_3,b_1,b_2,b_3\} $ are the coefficients of an affine transformation. The affine coefficients from all lenslets can reconstruct the dual-orthogonal coefficients $ {{e}_{p,q}} $ or discrete $ \phi \left( x_m,y_n,\xi_p ,\eta_q \right) $ (details in Appendix).

Various approaches, such as intensity-based optimization, control points, and feature detection and matching, can be used to register images on many platforms. In this paper, the MATLAB® internal function ‘imregtform’ is used to estimate affine transformation coefficients, in which the image similarity metric is optimized during registration.

\section{Numerical simulations}
\label{sec:simulations}

In this section, the numerical simulations in space-variant Shack-Hartmann wavefront sensing are given as the simulation configuration, the forward process of image formation, and wavefront reconstruction based on zonal and modal methods.

\subsection{Configuration and parameters for simulating space-variant Shack-Hartmann sensing}
\label{subsec:simulation configuration}

In this subsection, we set up the environments for simulating space-variant Shack-Hartmann sensing and wavefront reconstruction, and this configuration and parameters settled here are mainly consistent with the experiments in Section \ref{sec:Experiments}.

In our simulations, the SH has a microlens array with a 15 $ \times $ 15 lenslet, while each lenslet has a 500 \textmu m pitch and 46.7 mm focal length. A 465 $ \times $ 465 pixel camera is located at the focal plane of the microlens array. For simplicity, every lenslet covers 31 $ \times $ 31 pixels, while the pixel size is set at 500/31 \textmu m. The wavelength of the incident light is 462 nm. Additionally, a 1.2 $ \times $ 1.2 mm $ ^2 $ square light source lies on the front focal plane of a lens with a focal length of 200 mm, while the microlens array is located on its back focal plane, as shown in Fig. \ref{fig:Geometrical projection}(a). In addition, a 3 $ \times $ 3 star array source with a interval of 0.5 mm is used as a reference for comparison with our method, while the central star coincides with the center of the square source.

In order to generate the SH images of the square and the star array source, the measurement matrix of the SH system is calculated in advance. First, the space-variant wavefront is designed by using the geometrical projection method with different incident angles, where the variation of a wavefront is neglected during propagation from the phase screen to the lenslet array. The phase screen has a smooth distribution of PV 28.846 rad and RMS 5.712 rad, which are placed above the microlens array of nearly 195 mm, as shown in Fig. \ref{fig:Geometrical projection}.

The effect of a lenslet and the propagation from the lenslet array to the camera is considered as the Fourier transform property of a lens \cite{Goodman2017}. Thus, the fast Fourier transform (FFT) is implemented to obtain the complex amplitude of subimages on the corresponding local camera area. Thus, the whole camera image is assembled by all the subimages formed independently in each lenslet. By using this method, every wavefront, on behalf of one point on the object plane, produces a Shack-Hartmann point spread function (SH PSF), as shown in Fig. \ref{fig:Geometrical projection} (e). Then, all SH PSFs are combined as the measurement matrix. Finally, the SH images of the square and the star array source can be acquired by matrix multiplication. This process is almost identical to that used in light-field microscopy \cite{Broxton2013}. Before the vectorization, the object matrix is a 37 $ \times $ 37 matrix distributed by sampling the light source. The simulations are programmed in MATLAB® R2021a and run on a PC on the primary configurations of Intel i7-8700 CPU, 16G RAM with the Windows 10 professional operating system. The function ‘imregtform’, we used to obtain affine transform coefficients, is tested with many light sources, such as squares, circles, ellipses, octagons, and peaks (a function for producing continuous and smooth distribution by the MATLAB® internal function ‘peaks’) by simulations, and the results present consistent performance. This algorithm shows robustness for our applications. (details in Supplemental document). 

In space-variant wavefront reconstruction, the dual Legendre polynomials are naturally adopted in the modal method for the dual-orthogonal wavefront representation and reconstruction since the wavefront area and the object field are both settled squares in the simulations. Here, the two-dimensional Legendre polynomial formula is presented as

\begin{equation}
 {{L}_{m,n}}\left( x,y \right)=\frac{\text{1}}{{{\text{2}}^{m+n}}m!n!}\frac{{{d}^{m}}}{d{{x}^{m}}} {{\left( {{x}^{2}}-1 \right)}^{m}} \frac{{{d}^{n}}}{d{{y}^{n}}} {{\left( {{y}^{2}}-1 \right)}^{n}} ,
\label{eq:dual Legendre}
\end{equation}
where $ \left( x,y \right) $ are the lateral coordinates defined in the unit square [-1,1] $ ^2 $ and $ \left( m,n \right) $ the corresponding order. For mathematical convenience, the single index $ k $ is used to replace the double order $ \left( m,n \right) $ by the following rule:
\begin{equation}
 k=\frac{\left( m+n \right)\left( m+n+1 \right)}{2}+n,
\label{eq:dual Legendre order}
\end{equation}
that is,
\begin{equation}
 {{L}_{k}}\left( x,y \right)={{L}_{m,n}}\left( x,y \right).
\label{eq:dual Legendre 2}
\end{equation}

The total reconstruction term on the pupil plane is set at 45, namely, $P = 45$ in \eqref{eq:dual-orthogonal space-variant wavefront}, to represent the wavefront function since the spatial sampling is sufficient where the lenslet numbers are up to 15 $ \times $ 15. On the other hand, only 3 terms, also $Q = 3$ in \eqref{eq:dual-orthogonal space-variant wavefront}, are used and reconstructed for representing the space-variant function on the image plane because the affine transformation only characterizes very slow changes.

\subsection{Simulated results of the Shack-Hartmann sensor, space-variant function estimation, and wavefront reconstruction}
\label{subsec:Simulated images}
\begin{figure}[htbp!]
\centering
\includegraphics[width=\linewidth]{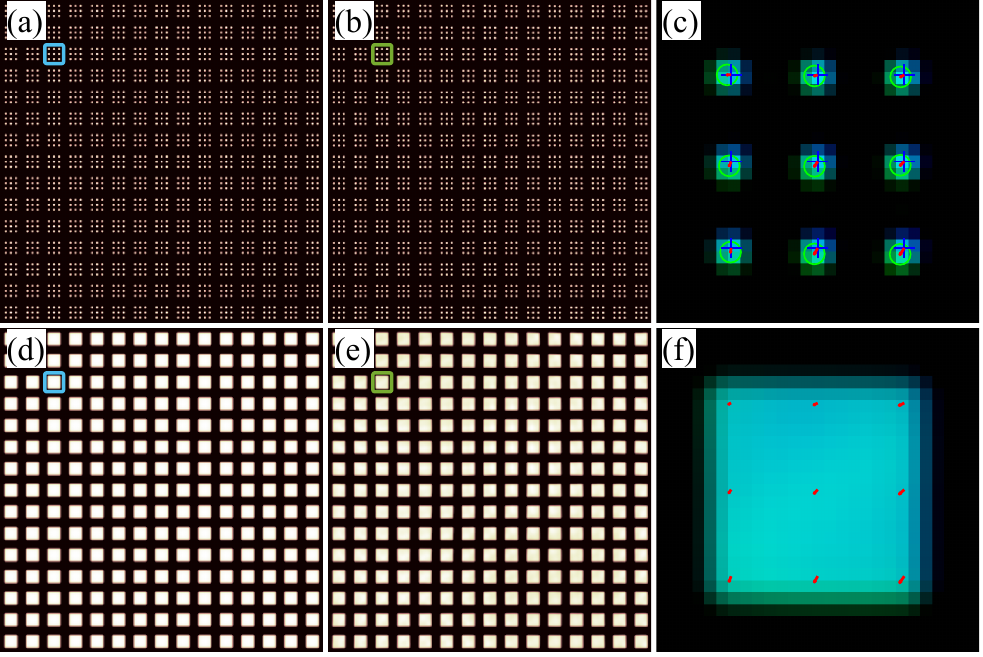}
\caption{Shack-Hartmann simulation and space-variant function estimation. The SH images are illuminated by the 3 $ \times $ 3 star array (a) without and (b) with the phase screen. (c) Subimages zoom of sections in the box of (a) and (b) labeled blue and green, centroid shift marked for each star. SH images illuminated by the square extended source (d) without and (e) with phase screen. (f) Subimages zoom of sections in the box of (d) and (e) labeled blue and green, position shift marked by the affine transform estimation.}
\label{fig:Simulated images}
\end{figure}

By using the forward process algorithm described in Section \ref{subsec:simulation configuration} and Fig. \ref{fig:Geometrical projection}, the simulation results of SH images are shown in Fig. \ref{fig:Simulated images} under the square and point array illumination sources with and without phase perturbation. 

Two subimages with and without the phase screen are marked blue and green, respectively, as shown in Fig. \ref{fig:Simulated images}(c, f). For the point array source, the centroids of all subspots are extracted conventionally to calculate centroid shifts, . For the square source, the centroid shifts are calculated at the blue points in Fig. \ref{fig:Simulated images}(c) based on the space-variant function obtained by affine transform estimation, as shown in Fig. \ref{fig:Simulated images}(f).

\begin{figure}[ht]
\centering
\includegraphics[width=\linewidth]{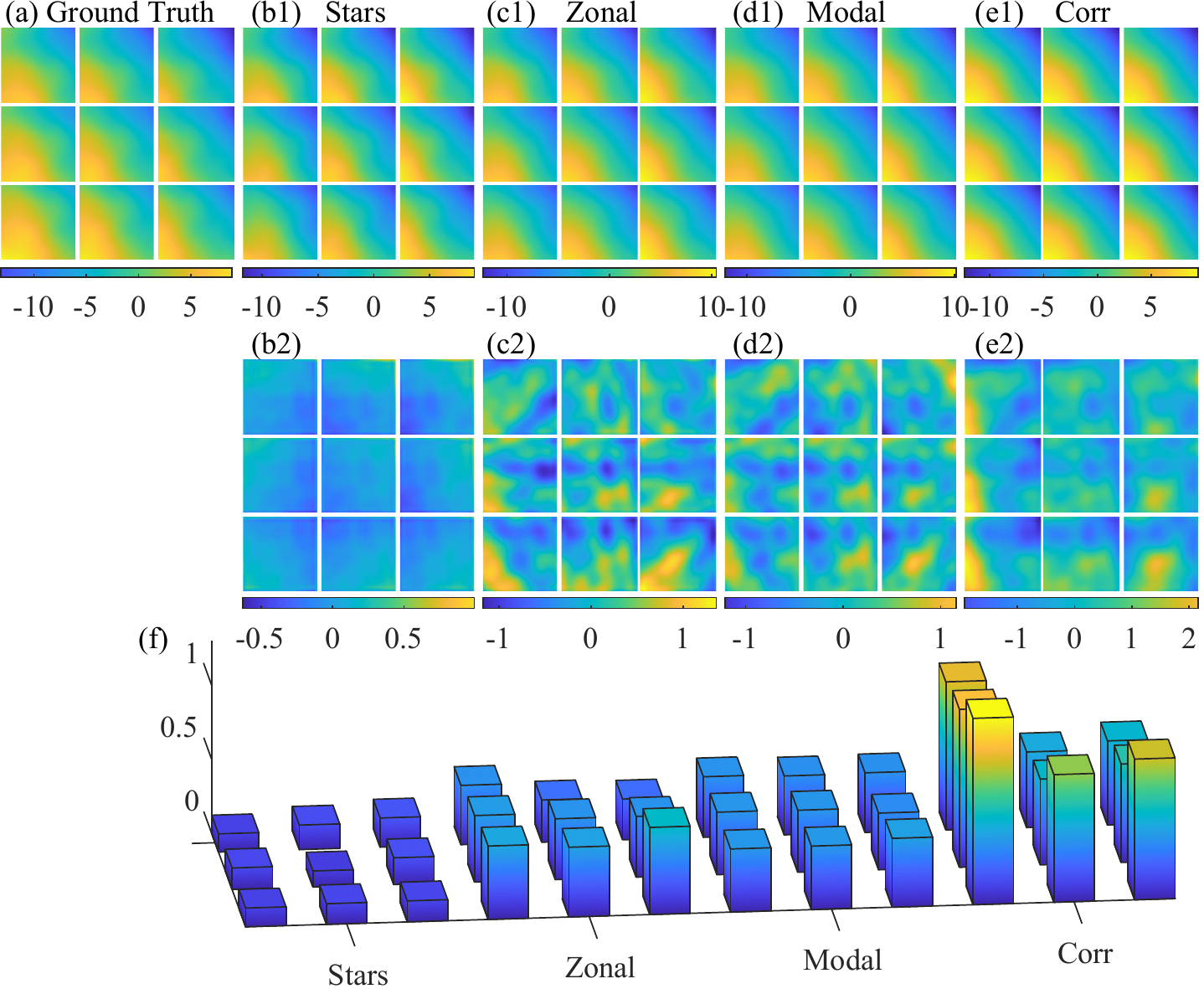}
\caption{Comparisons of simulation results of wavefront reconstruction, all scaled in radians. (a) Projected wavefronts with different angles of view, the ground truth. Reconstruction wavefronts by using (b1) the zonal method with 3 $ \times $ 3 stars illuminated, proposed (c1) zonal and (d1) modal methods, and (e1) the correlation method with square source illuminated. (b2-e2) Residual wavefronts between (a) the ground truth and (b1-e1) the corresponding reconstruction wavefronts. (f) RMS of the corresponding residual wavefronts (b2-e2).}
\label{fig:Simulation results}
\end{figure}

For the zonal method, the corresponding wavefront can be estimated based on the Southwell approach \cite{Southwell1980} once we have the centroid shifts or the local slopes, also known as the zonal method. The wavefront reconstruction results under 3 $ \times $ 3 point array illumination are provided in Fig. \ref{fig:Simulation results}(b1) as well as the residual wavefronts in Fig. \ref{fig:Simulation results}(b2). The RMS of the residual wavefronts varies from 0.108 to 0.197 rad across different views, whereas that of the input wavefront varies from 3.681 to 4.765 rad. The reconstructed wavefronts by the zonal method under square hole illumination are presented in Fig. \ref{fig:Simulation results}(c1), as well as the residual wavefronts in Fig. \ref{fig:Simulation results}(c2). These RMS values of the residual wavefronts vary from 0.275 to 0.589 rad.
For the modal method, the reconstructed wavefronts are obtained and presented in Fig. \ref{fig:Simulation results}(d1), as well as the residual wavefronts in Fig. \ref{fig:Simulation results}(d2). The RMS of the residual wavefronts varies from 0.383 to 0.465 rad.

In addition, the correlation method for conventional space-invariant Shack-Hartmann wavefront sensing is carried out with an extended light source for comparison. In this method, the correlation algorithm is performed between the subimages with and without aberration to evaluate the local shift. The wavefronts reconstructed by the correlation method are presented in Fig. \ref{fig:Simulation results}(e1) as well as the residual wavefronts in Fig. \ref{fig:Simulation results}(e2). The RMS of the residual wavefronts varies from 0.511 to 1.260 rad.

In summary, the RMS of the residual is presented in Tab.\ref{tab:simulation}. The 3 $ \times $ 3 stars provide the best reconstruction results naturally; the extended light source with our space-variant approach is the next while the zonal and modal methods have a similar performance. The modal method provides more averaged results than the zonal method. Notably, for the zonal method, the maximal RMS is nearly twice as much as the minimal RMS. The conventional correlation method has the largest RMS since it does not consider the space-variant modal. In addition, the reconstruction of the central view usually outperforms that of the periphery views.

\begin{table}[ht]
\centering
\caption{Comparison of the RMS of residual wavefronts referring to different cases in numerical simulations, all scaled in radians.}
\begin{tabular}{ccccc}
\hline
 & Star & Zonal & Modal & Corr \\
\hline
  Minimal RMS & 0.108 & 0.275 & 0.383 & 0.511 \\ 
  Maximal RMS & 0.197 & 0.589 & 0.465 & 1.260 \\ 
  Mean RMS & 0.147 & 0.421 & 0.417 & 0.830 \\ 
\hline
\end{tabular}
 \label{tab:simulation}
\end{table}

\section{Experiments}
\label{sec:Experiments}
In this section, an experimental demonstration is implemented to validate our proposal for space-variant Shack-Hartmann wavefront sensing. The configuration of the experiments and wavefront reconstruction results are presented.
\subsection{Experimental setup}
\label{subsec:Experimental setup}
\begin{figure}[htbp!]
\centering
\includegraphics[width=\linewidth]{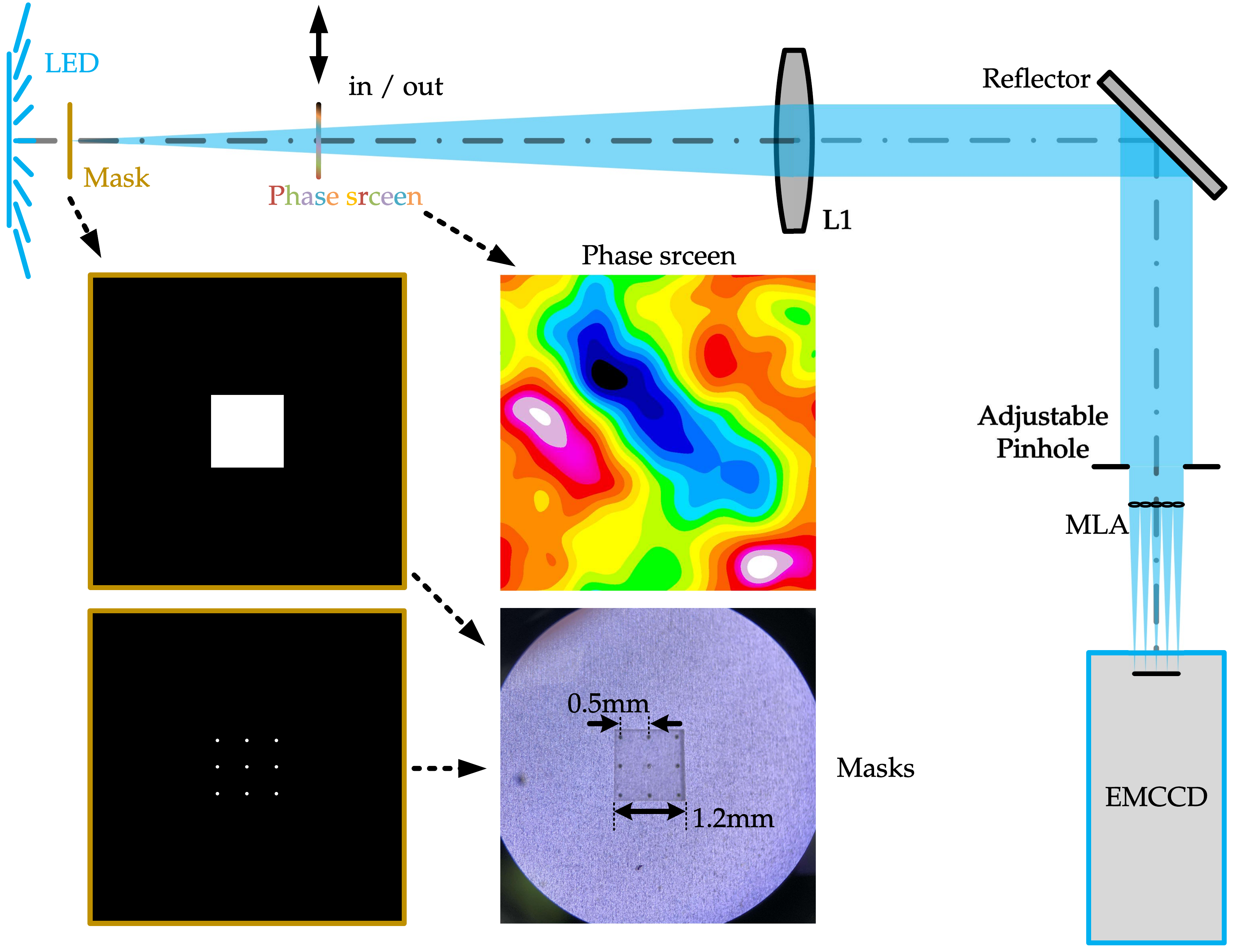}
\caption{Schematic of the main experimental setup for space-variant Shack-Hartmann wavefront sensing. L1: 200 mm lens, MLA: microlens array, EMCCD: electron-multiplying CCD camera.}
\label{fig:Experimental setup}
\end{figure}

As shown in Fig. \ref{fig:Experimental setup}, we build the demonstration system for space-variant Shack-Hartmann wavefront sensing. To perform the designated light source, an LED diffuser and a mask are combined, in which the LED with a center wavelength of 462 nm provides a large area of illumination with a broad illumination field of view and the mask shapes the images on the camera. Hence, two customized masks are used, including a 1.2 $ \times $ 1.2 mm $ ^2 $ square for performing the extended light source and a 3 $ \times $ 3 pinhole array for reference, where each pinhole has a 50 \textmu m diameter and intervals of 0.5 mm. The image of the two masks overlapped is also shown in Fig. \ref{fig:Experimental setup}. The mask is located on the front focal plane of lens L1 with a focal length of 200 mm, and the LED diffuser and phase screen lie above and below the mask by approximately 10 mm and 20 mm, respectively. The light beam is collimated by the lens L1 and projected onto a homemade Shack-Hartmann wavefront sensor. The SH is made by a microlens array (Edmund, $\#$64-483, pitch 0.5 mm, focal length 46.7 mm, size 10 $ \times $ 10 mm $ ^2 $ ) and an electron-multiplying CCD (Hamamatsu, C9100-23B, 512 $ \times $ 512 pixels, cell size 16 $ \times $ 16 \textmu m $ ^2 $ ), and the camera is located on the back focal plane of the microlens array. The exposure times are set at 3.8 s and 150 ms to acquire images to the pinhole array and square mask, respectively, while the normal mode of CCD is used instead of the multiplying mode. The phase screen is made by smearing epoxy glue on a glass slide.

\subsection{Experimental results of the Shack-Hartmann sensor, space-variant function estimation, and wavefront reconstruction}
\label{subsec:Experimental images}
\begin{figure}[htbp!]
\centering
\includegraphics[width=\linewidth]{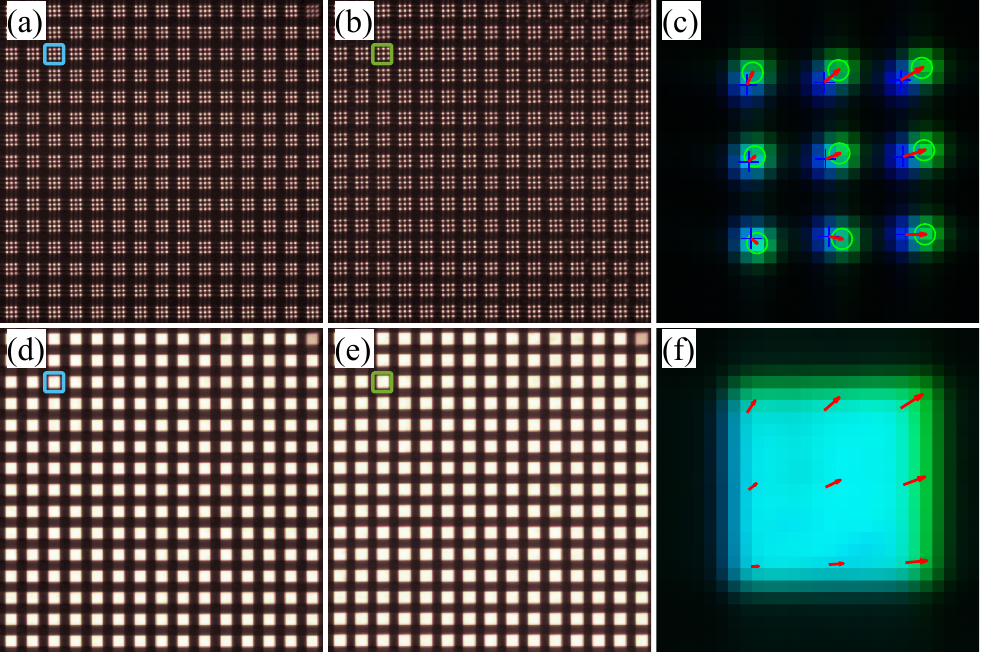}
\caption{Acquired Shack-Hartmann images and space-variant function estimation. The SH images are illuminated by the 3 $ \times $ 3 star array (a) without and (b) with the phase screen. (c) Subimages zoom of sections in the box of (a) and (b) labeled blue and green, centroid shift marked for each star. SH images illuminated by the square extended source (d) without and (e) with phase screen. (f) Subimages zoom of sections in the box of (d) and (e) labeled blue and green, position shift marked by the affine transform estimation.}
\label{fig:Experimental images}
\end{figure}

The SH images acquired by our experimental setup are shown in Fig. \ref{fig:Experimental images}. The periphery pixels of raw 512 $ \times $ 512 images were chopped to 469 $ \times $ 469 images and then resized to 465 $ \times $ 465 based on the bicubic interpolation method for nearly 15 $ \times $ 15 pixels in each lenslet. This processing is implemented to conveniently extract the subimages from each lenslet. Thus, the affine transform estimation is used to provide the subtle shifts compared to the reference.

\begin{figure}[ht]
\centering
\includegraphics[width=\linewidth]{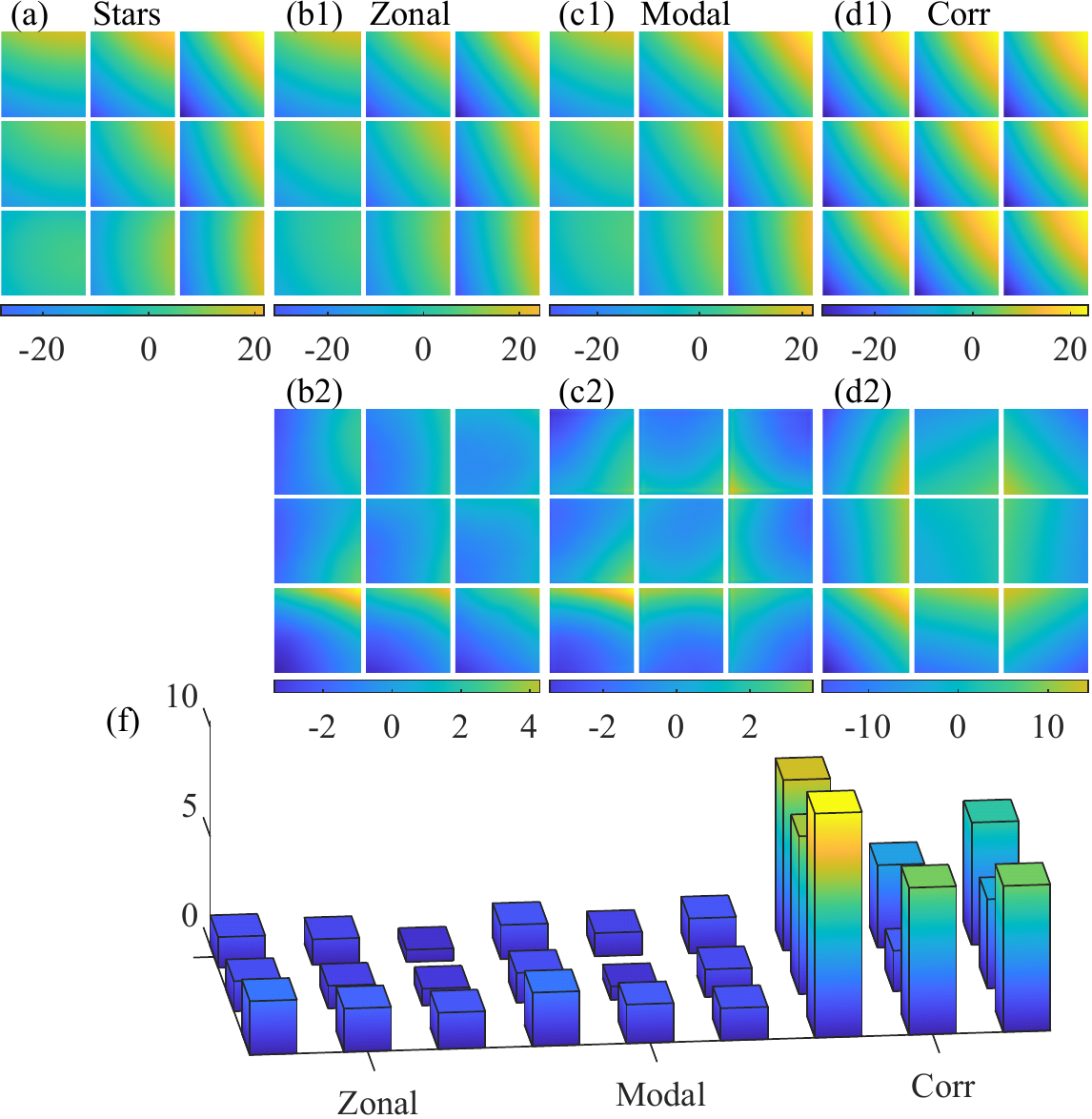}
\caption{Comparisons of experimental results of wavefront reconstruction, all scaled in radians. Reconstruction wavefronts by using (a) the zonal method with 3 $ \times $ 3 stars illuminated, proposed approach with (b1) zonal and (c1) modal methods, and (d1) the correlation method with square source illuminated. (b2-d2) Residual wavefronts between the (a) reference results and (b1-d1) the corresponding reconstruction wavefronts. (f) RMS of the corresponding residual wavefronts (b2-d2).}
\label{fig:Experimental results}
\end{figure}

The reconstruction results of wavefronts are shown in Fig. \ref{fig:Experimental results}(a), and the RMS of residual wavefronts is summarized in Tab.\ref{tab:Experimental}. The reconstructed wavefronts under star array illumination exhibit obvious and continuously slow space-variant features. The reconstructed wavefronts under square hole illumination are presented in Fig. \ref{fig:Experimental results}(b1-d1) via our zonal and modal method as well as the correlation method. By using the wavefronts of star array illumination as a reference, the corresponding residual wavefronts are obtained and shown in Fig. \ref{fig:Experimental results}(b2-d2). The RMS of the residual wavefronts via our zonal method varies from 0.558 to 2.450 rad. The modal method provides close results that vary from 0.617 to 2.449 rad; thus, the zonal and modal methods used in our approach present similar performance in reconstructing wavefronts. The correlation method shows that the RMS varies from 1.850 to 10.228 rad. Further comparisons could be exploited with different RMSs or PVs of wavefront and space-variant functions in the future.

\begin{table}[ht]
\centering
\caption{Comparison of the RMS of residual wavefronts referring to different cases in imaging experiments, all scaled in radians.}
\begin{tabular}{cccc}
\hline
 & Zonal & Modal & Corr \\
\hline
  Minimal RMS & 0.558 & 0.617 & 1.850 \\ 
  Maximal RMS & 2.450 & 2.449 & 10.228 \\ 
  Mean RMS & 1.378 & 1.461 & 5.984 \\ 
\hline
\end{tabular}
 \label{tab:Experimental}
\end{table}

\section{Discussion}
\label{sec:Discussion}
In the experiments, the absolute values of RMS of residual wavefronts are still high via our methods, as shown in Tab.\ref{tab:Experimental}. This phenomenon is mainly because the amplitude of the input phase is significant; correspondingly, its high-frequency components are also high. In addition, point array illumination reveals the spatially local wavefront, while the square source with affine transformation only carried out the spatially averaged global wavefront. Thus, the spatially high-frequency components may not be well reconstructed within the square source, resulting in a high absolute value of residual wavefronts RMS.

Another interesting result is that the reconstructed wavefront at any FOV did not represent much of the high-frequency component in the back pupil plane, as shown in \ref{fig:Experimental results}(a). There are three main contributions to these results. The first one is the epoxy glue-made phase screen, while the strong tension of the glue made its surface very smooth. In addition, every point of the extended light source only bites a small piece on the screen with a very narrow angle since the source has a close distance to the phase screen. Moreover, the transformation from the screen plane to the back focal plane, including two free-space propagations and a lens, further smooths this partial phase.

It is worth mentioning that the optical flow method in \cite{Wu2021} and our affine transformation estimation method could be classified into the same category. The traditional light field camera is essentially equivalent to the SH with the difference that the microlens is located on the imaging plane or the pupil plane \cite{Lam2015,Ko2017}. After transferring the light-field camera data from the spatial domain to the phase-space domain by corresponding pixel rearrangement, each light-field subimage is conjugated to the object with local wavefront perturbations at every object point due to the space-variant wavefront, as in SH. In phase space, light-field images and SH images all present so-called instantaneous phase information. With point light source illumination, the instantaneous phase reveals the gradient of the wavefront phase and thus the phase itself. With the extended light source illumination, the optical flow method is implemented by searching the maximum correlation positions for 7 $ \times $ 7 segmented areas and then using cubic interpolation to obtain the shifts of every pixel. The same a priori assumption is used here that aberrations change slowly across the whole FOV. In our proposal, we designed a uniform extended guide source that tiled on the entire FOV, but the optical flow method directly used the distribution of the objective block by a block that could be unevenly distributed. The uniformity of subimages may affect the accuracy of wavefront reconstruction. Additionally, the optimization-based algorithm in affine transformation estimation generally outperforms the blocked correlation methods and cubic interpolation.

Furthermore, the affine transformation can only characterize the low-frequency deformation of the subimages. If the sample changes rapidly, our method may lose fidelity. In these cases, other nonrigid intensity-based techniques, such as diffeomorphic demons \cite{Vercauteren2009} and free form deformation \cite{Xie2004}, could provide the localized displacement field with more high-frequency components. However, with no modal characterization, the modal method and representation may not feasible to the displacement field directly.

\section{Conclusion}
\label{sec:Conclusion}
In this paper, we propose the space-variant Shack-Hartmann wavefront sensing approach based on affine transformation estimation. Two main contributions are made. First, the affine transformation is adopted to estimate local shifts at every FOV under extended source illumination since the conventional correlation estimation only expresses the overall displacement with no local details. Second, we provide the dual-orthogonal model for space-variant wavefront representation and build a framework of wavefront reconstruction with this model. For comparison, the space-variant wavefront was previously represented by many wavefronts distributed at different FOVs with no efficiency and no elegance.

We carried out simulations and experiments on the space-variant Shack-Hartmann wavefront sensing approach. In the simulations, the average RMS of 0.83 rad of the resident wavefront reconstructed by using the conventional space-invariant Shack-Hartmann correlation method is reduced to 0.42 rad by using our methods, showing double performance improvements. In the experiments, the average RMS 6 rad of the correlation method is reduced to 1.4 rad of ours with quadruple improvements. The results clearly support our theory and method.

\section{Appendix}
\label{sec:Appendix}
We review the formulas in the space-invariant case of how the Shack-Hartmann sensor detects wavefront and then generalize them into the space-variant case.

Hence, let $ {{\phi }^{m,n}}\left( {{\xi }_{L}},{{\eta }_{L}} \right) $ represents the local wavefront of the global wavefront $ \phi \left( \xi ,\eta \right) $ at the $ m  $th row and $ n  $th column of the lenslet array with $ M\times N $ subapertures, where $ \left( {{\xi }_{L}},{{\eta }_{L}} \right) $ is the local lateral coordinates while its origin settles on the center of the microlens. Since the global wavefront is slowly varying, the local wavefront is considered a local slope

\begin{equation}
 {{\phi }^{m,n}}\left( {{\xi }_{L}},{{\eta }_{L}} \right)={{a}^{m,n}}{{\xi }_{L}}+{{b}^{m,n}}{{\eta }_{L}}+{{t}^{m,n}},
\label{eq:local slope-A}
\end{equation}
where $ {{a}^{m,n}} $ , $ {{b}^{m,n}} $ , and $ {{t}^{m,n}} $ are arbitrary coefficients. Thus, the local gradients of the global wavefront are represented by spot positions \cite{Feng2018}

\begin{equation}
 \begin{aligned}
  & A \left( x_{2}^{m,n}-x_{1}^{m,n} \right)=B{{a}^{m,n}}, \\ 
  & A \left( y_{2}^{m,n}-y_{1}^{m,n} \right)=B{{b}^{m,n}}, \\ 
 \end{aligned}
\label{eq:local gradients-A}
\end{equation}
where $ \left( x_{1}^{m,n},y_{1}^{m,n} \right) $ and $ \left( x_{2}^{m,n},y_{2}^{m,n} \right) $ are the normalized coordinates of a arbitrary point in $ (m,n)  $th lenslet without and with aberration, respectively, and 
 $ A = D/2 $ , which transfers the normalized centroid shift to actual physical shift, $ D $ is the pitch of the lenslet, and $ B = f/k $ , $ f $ is the focal length of the lenslet and $ k $ is the wavenumber.

In the case of space-variant wavefront sensing, the local slope varies in different positions

\begin{equation}
	{{\phi }^{m,n}}\left( x,y,{{\xi }_{L}},{{\eta }_{L}} \right)={{a}^{m,n}}\left( x,y \right){{\xi }_{L}}+{{b}^{m,n}}\left( x,y \right){{\eta }_{L}}+{{t}^{m,n}}\left( x,y \right),
\label{eq:space-variant local slope function-A}
\end{equation}
where $ {{a}^{m,n}}\left( x,y \right) $ , $ {{b}^{m,n}}\left( x,y \right) $ , and $ {{t}^{m,n}}\left( x,y \right) $ are the coefficient functions along with object coordinates. Then, the global wavefront is reconstructed from the map of the coefficient functions by zonal or modal representation.

Assuming the space-variant function is moderate enough, these gradients can be modeled by a slope function
\begin{equation}
\begin{aligned}
 {{a}^{m,n}}\left( x,y \right)&=a_{1}^{m,n}x+a_{2}^{m,n}y+a_{3}^{m,n}, \\ 
 {{b}^{m,n}}\left( x,y \right)&=b_{1}^{m,n}x+b_{2}^{m,n}y+b_{3}^{m,n}, \\ 
\end{aligned}
\label{eq:space-variant gradients-A}
\end{equation}
while the piston $ {{t}^{m,n}}\left( x,y \right) $ cannot be represented in the gradients or subimages, and they can be reconstructed in the zonal or modal method under the continuous condition of the global wavefront. Thus, the local slope drives the arbitrary point $ \left( x_{1}^{m,n},y_{1}^{m,n} \right) $ to $ \left( x_{2}^{m,n},y_{2}^{m,n} \right) $ with the relation

\begin{equation}
\left[ {\begin{array}{*{20}{c}}
{x_2^{m,n} - x_1^{m,n}}\\
{y_2^{m,n} - y_1^{m,n}}\\
1
\end{array}} \right] = \left[ {\begin{array}{*{20}{c}}
{a_1^{m,n}}&{a_2^{m,n}}&{a_3^{m,n}}\\
{b_1^{m,n}}&{b_1^{m,n}}&{b_1^{m,n}}\\
0&0&1
\end{array}} \right]\left[ {\begin{array}{*{20}{c}}
{x_1^{m,n}}\\
{y_1^{m,n}}\\
1
\end{array}} \right],
\label{eq:centroid and slope-A}
\end{equation}
or
\begin{equation}
\left[ {\begin{array}{*{20}{c}}
{x_2^{m,n}}\\
{y_2^{m,n}}\\
1
\end{array}} \right] = \left[ {\begin{array}{*{20}{c}}
{a_1^{m,n} + 1}&{a_2^{m,n}}&{a_3^{m,n}}\\
{b_1^{m,n}}&{b_1^{m,n} + 1}&{b_1^{m,n}}\\
0&0&1
\end{array}} \right]\left[ {\begin{array}{*{20}{c}}
{x_1^{m,n}}\\
{y_1^{m,n}}\\
1
\end{array}} \right],
\label{eq:centroid and slope2-A}
\end{equation}
where the coefficients $ A $ and $ B $ in \eqref{eq:local gradients-A} are concealed for mathematical simplicity. As shown in Fig. \ref{fig:Schematic}(c2), the coefficients in \eqref{eq:centroid and slope2-A} are characterized and estimated by an affine transformation estimation between the subimages with and without the aberration. 

The wavefront is about to be reconstructed from the map of the estimated local information obtained via zonal and modal representation based on the smooth \textit{prior} of the global wavefront. The zonal method assumes that the phases on the edges of the neighboring lenslets are equal, and the modal method decomposes the wavefront into a sum of a series of continuous orthogonal polynomials.

For the zonal method, the local slope $ {{a}^{m,n}} $ , $ {{b}^{m,n}} $ of any point $ \left( {{x}_{0}},{{y}_{0}} \right) $ on the object plane can be obtained by the local slope function in \eqref{eq:space-variant local slope function-A} based on the result of the affine transformation estimation. Then, the space-variant wavefront $ {{\phi }^{m,n}}\left( {{x}_{0}},{{y}_{0}} \right) $ can be reconstructed from the map of the local slope $ {{a}^{m,n}} $ , $ {{b}^{m,n}} $ by using the Southwell approach \cite{Southwell1980}.

For the modal method, a proper dimensionality-reduced presentation is required to reconstruct the wavefront by the modal method effectively. Through several form transformations with \eqref{eq:dual-orthogonal space-variant wavefront} and \eqref{eq:centroid and slope-A}, the relation between the dual-orthogonal coefficients $ {{e}_{p,q}} $ and the local slope function coefficients $ a_{i}^{m,n},b_{i}^{m,n},i=1,2,3 $ is given as

\begin{equation}
\begin{aligned}
 & a_{1}^{m,n}=\sum\limits_{q=1}^{Q}{\sum\limits_{p=1}^{P}{{{e}_{p,q}}\frac{\partial {{T}_{q}}\left( x,y \right)}{\partial x}\frac{\partial {{R}_{p}}\left( \xi ,\eta \right)}{\partial \xi }}}, \\ 
 & a_{2}^{m,n}=\sum\limits_{q=1}^{Q}{\sum\limits_{p=1}^{P}{{{e}_{p,q}}\frac{\partial {{T}_{q}}\left( x,y \right)}{\partial y}\frac{\partial {{R}_{p}}\left( \xi ,\eta \right)}{\partial \xi }}}, \\ 
 & a_{3}^{m,n}=\sum\limits_{p=1}^{P}{{{e}_{p,1}}\frac{\partial {{R}_{p}}\left( \xi ,\eta \right)}{\partial \xi }}, \\ 
 & b_{1}^{m,n}=\sum\limits_{q=1}^{Q}{\sum\limits_{p=1}^{P}{{{e}_{p,q}}\frac{\partial {{T}_{q}}\left( x,y \right)}{\partial x}\frac{\partial {{R}_{p}}\left( \xi ,\eta \right)}{\partial \eta }}}, \\ 
 & b_{2}^{m,n}=\sum\limits_{q=1}^{Q}{\sum\limits_{p=1}^{P}{{{e}_{p,q}}\frac{\partial {{T}_{q}}\left( x,y \right)}{\partial y}\frac{\partial {{R}_{p}}\left( \xi ,\eta \right)}{\partial \eta }}}, \\ 
 & b_{3}^{m,n}=\sum\limits_{p=1}^{P}{{{e}_{p,1}}\frac{\partial {{R}_{p}}\left( \xi ,\eta \right)}{\partial \eta }}. \\ 
\end{aligned}
\label{eq:coefficients and coefficients relation-A}
\end{equation}
Thus, \eqref{eq:coefficients and coefficients relation-A} can be written as follows:
\begin{equation}
\mathbf{d}=\mathbf{Me},
\label{eq:Matrix-A}
\end{equation}
where
\[{\bf{d}} = {\left[ {\begin{array}{*{20}{c}}
{{\bf{a}}_1^{m,n}}&{{\bf{a}}_2^{m,n}}&{{\bf{a}}_3^{m,n}}&{{\bf{b}}_1^{m,n}}&{{\bf{b}}_2^{m,n}}&{{\bf{b}}_3^{m,n}}
\end{array}} \right]^T,}\]

\[\begin{array}{l}
{\bf{a}}_1^{m,n} = {\left[ {\begin{array}{*{20}{c}}
{a_1^{1,1}}&{a_1^{1,2}}& \ldots &{a_1^{M,N}}
\end{array}} \right]},\\
{\bf{a}}_2^{m,n} = {\left[ {\begin{array}{*{20}{c}}
{a_2^{1,1}}&{a_2^{1,2}}& \ldots &{a_2^{M,N}}
\end{array}} \right]},\\
{\bf{a}}_3^{m,n} = {\left[ {\begin{array}{*{20}{c}}
{a_3^{1,1}}&{a_3^{1,2}}& \ldots &{a_3^{M,N}}
\end{array}} \right]},\\
{\bf{b}}_1^{m,n} = {\left[ {\begin{array}{*{20}{c}}
{b_1^{1,1}}&{b_1^{1,2}}& \ldots &{b_1^{M,N}}
\end{array}} \right]},\\
{\bf{b}}_2^{m,n} = {\left[ {\begin{array}{*{20}{c}}
{b_2^{1,1}}&{b_2^{1,2}}& \ldots &{b_2^{M,N}}
\end{array}} \right]},\\
{\bf{b}}_3^{m,n} = {\left[ {\begin{array}{*{20}{c}}
{b_3^{1,1}}&{b_3^{1,2}}& \ldots &{b_3^{M,N}}
\end{array}} \right],}
\end{array}\]

\begin{equation*}
\begin{aligned}
 & \mathbf{e}={{\left[ \begin{matrix}
 {{\mathbf{e}}_{p,1}} & {{\mathbf{e}}_{p,2}} & \cdots & {{\mathbf{e}}_{p,q}} \\
\end{matrix} \right]}^{T}} \\ 
 & ={{\left[ \begin{matrix}
 {{e}_{1,1}} & \ldots & {{e}_{p,1}} & {{e}_{1,2}} & \ldots & {{e}_{p,2}} & \ldots & {{e}_{1,q}} & \ldots & {{e}_{p,q}} \\
\end{matrix} \right]}^{T},}
\end{aligned}
\end{equation*}

\begin{small}
\[{\bf{M}} = \left[ {\begin{array}{*{20}{c}}
{\frac{{\partial {T_1}\left( {x,y} \right)}}{{\partial x}}\frac{{\partial {{\bf{R}}_p}\left( {\xi ,\eta } \right)}}{{\partial \xi }}}&{\frac{{\partial {T_2}\left( {x,y} \right)}}{{\partial x}}\frac{{\partial {{\bf{R}}_p}\left( {\xi ,\eta } \right)}}{{\partial \xi }}}& \cdots &{\frac{{\partial {T_Q}\left( {x,y} \right)}}{{\partial x}}\frac{{\partial {{\bf{R}}_p}\left( {\xi ,\eta } \right)}}{{\partial \xi }}}\\
{\frac{{\partial {T_1}\left( {x,y} \right)}}{{\partial y}}\frac{{\partial {{\bf{R}}_p}\left( {\xi ,\eta } \right)}}{{\partial \xi }}}&{\frac{{\partial {T_2}\left( {x,y} \right)}}{{\partial y}}\frac{{\partial {{\bf{R}}_p}\left( {\xi ,\eta } \right)}}{{\partial \xi }}}& \cdots &{\frac{{\partial {T_Q}\left( {x,y} \right)}}{{\partial y}}\frac{{\partial {{\bf{R}}_p}\left( {\xi ,\eta } \right)}}{{\partial \xi }}}\\
{\frac{{\partial {{\bf{R}}_p}\left( {\xi ,\eta } \right)}}{{\partial \xi }}}&0& \cdots &0\\
{\frac{{\partial {T_1}\left( {x,y} \right)}}{{\partial x}}\frac{{\partial {{\bf{R}}_p}\left( {\xi ,\eta } \right)}}{{\partial \eta }}}&{\frac{{\partial {T_2}\left( {x,y} \right)}}{{\partial x}}\frac{{\partial {{\bf{R}}_p}\left( {\xi ,\eta } \right)}}{{\partial \eta }}}& \cdots &{\frac{{\partial {T_Q}\left( {x,y} \right)}}{{\partial x}}\frac{{\partial {{\bf{R}}_p}\left( {\xi ,\eta } \right)}}{{\partial \eta }}}\\
{\frac{{\partial {T_1}\left( {x,y} \right)}}{{\partial y}}\frac{{\partial {{\bf{R}}_p}\left( {\xi ,\eta } \right)}}{{\partial \eta }}}&{\frac{{\partial {T_2}\left( {x,y} \right)}}{{\partial y}}\frac{{\partial {{\bf{R}}_p}\left( {\xi ,\eta } \right)}}{{\partial \eta }}}& \cdots &{\frac{{\partial {T_Q}\left( {x,y} \right)}}{{\partial y}}\frac{{\partial {{\bf{R}}_p}\left( {\xi ,\eta } \right)}}{{\partial \eta }}}\\
{\frac{{\partial {{\bf{R}}_p}\left( {\xi ,\eta } \right)}}{{\partial \eta }}}&0& \cdots &0
\end{array}} \right]\]
\end{small}

\[\frac{\partial {{\mathbf{R}}_{p}}\left( \xi ,\eta \right)}{\partial k}=\left[ \begin{matrix}
 \frac{\partial {{R}_{1}}\left( \xi ,\eta \right)}{\partial k} & \frac{\partial {{R}_{2}}\left( \xi ,\eta \right)}{\partial k} & \cdots & \frac{\partial {{R}_{P}}\left( \xi ,\eta \right)}{\partial k} \\
\end{matrix} \right],k=\left\{ \xi ,\eta \right\}.\]

Finally, the solution of \eqref{eq:Matrix-A} is given as
\begin{equation}
 \mathbf{e}=\text{pinv}\left( \mathbf{M} \right)\mathbf{d}
\label{eq:solution Matrix-A}
\end{equation}
where pinv means pseudoinverse operation.

\begin{backmatter}
\bmsection{Funding} National Natural Science Foundation of China (32071458, 21927813, 81827809, 81925022, 92054301, 92150301), Beijing Natural Science Foundation Key Research Topics (Z20J00059).

\bmsection{Disclosures} The authors declare that there are no conflicts of interest related to this article.

\bmsection{Data Availability Statement} Data underlying the results presented in this paper are not publicly available at this time but may be obtained from the authors upon reasonable request.

\end{backmatter}

% Bibliography
\bibliography{sample}

% Full bibliography added automatically for Optics Letters submissions; the following line will simply be ignored if submitting to other journals.
% Note that this extra page will not count against page length
%\bibliographyfullrefs{sample}

%Manual citation list
%\begin{thebibliography}{1}
%\bibitem{Zhang:14}
%Y.~Zhang, S.~Qiao, L.~Sun, Q.~W. Shi, W.~Huang, %L.~Li, and Z.~Yang,
 % \enquote{Photoinduced active terahertz metamaterials with nanostructured
 %vanadium dioxide film deposited by sol-gel method,} Opt. Express \textbf{22},
 %11070--11078 (2014).
%\end{thebibliography}

\end{document}